\documentclass[aps,pre,superscriptaddress,twocolumn]{revtex4-1}

\usepackage{graphicx}

\usepackage{amsmath}
\usepackage{amssymb}
\usepackage[outdir=./]{epstopdf}
\usepackage[dvipsnames]{xcolor}

\begin{document}

\title{Elastohydrodynamic lift at a soft wall}

\author{Heather S. Davies} 
\affiliation{Univ. Grenoble Alpes, CNRS, LIPhy, 38000 Grenoble, France}

\author{Delphine D{\'e}barre} 
\affiliation{Univ. Grenoble Alpes, CNRS, LIPhy, 38000 Grenoble, France}

\author{Nouha El Amri} 
\affiliation{Univ. Grenoble Alpes, CNRS, LIPhy, 38000 Grenoble, France}

\author{Claude Verdier}
\affiliation{Univ. Grenoble Alpes, CNRS, LIPhy, 38000 Grenoble, France}

\author{Ralf P. Richter}
\email[]{R.Richter@leeds.ac.uk}
\affiliation{School of Biomedical Sciences, Faculty of Biological Sciences, School of Physics and Astronomy, Faculty of Mathematics and Physical Sciences, Astbury Centre for Structural Molecular Biology, University of Leeds,
Leeds LS2 9JT, UK}
\affiliation{CIC biomaGUNE, Paseo Miramon 182, 20014 San Sebastian, Spain}

\author{Lionel Bureau}
\email[]{lionel.bureau@univ-grenoble-alpes.fr}
\affiliation{Univ. Grenoble Alpes, CNRS, LIPhy, 38000 Grenoble, France}

\date{\today}

\begin{abstract}
We study experimentally the motion of non-deformable microbeads in a linear shear flow close to a wall bearing a thin and soft polymer layer. Combining microfluidics and 3D optical tracking, we demonstrate that the steady-state bead/surface distance increases with the flow strength. Moreover, such lift is shown to result from flow-induced deformations of the layer, in quantitative agreement with theoretical predictions from elastohydrodynamics. This study thus provides the first experimental evidence of ``soft lubrication'' at play at small scale, in a system relevant {\it e.g.} to the physics of blood microcirculation.
\end{abstract}


\maketitle

Elastohydrodynamics (EHD) is a key concept in soft matter physics \cite{Wang:2017hn,Chan:2009ed,Wang:2015cm}. The coupling between flow-induced pressure fields and elasticity of immersed objects is at the heart of topics ranging from the rheology of soft colloids \cite{Vlassopoulos:2014iv} to microfluidic particle sorting \cite{Geislinger:2014gu} and contact-free mechanical probe techniques \cite{Leroy:2012dy}. EHD is also central to biophysical problems, such as swimming of micro-organisms \cite{Goldstein:2016bm}, lubrication in synovial joints and blood microcirculation \cite{Jin:2006bx}. In the latter context, EHD governs the radial migration of circulating blood cells, which underlies vascular processes, such as margination \cite{Geislinger:2014gu,Kumar:2012ie,Zhao:2012gg}: leukocytes and platelets flow preferentially close to the vessel walls, while softer red blood cells (RBCs) migrate away from them. This gives rise to a cell-free layer, a $\mu$m-thick region forming near the vascular walls and depleted of RBCs \cite{Kim:2009jf}. This has been characterized {\it in vitro}, through flow experiments studying how RBCs \cite{Grandchamp:2013jq} or model vesicles \cite{Abkarian:2002ky} are repelled by a surface. The classical interpretation for the formation of the cell-free layer is that RBCs flowing near a surface deform under the fluid shear stress and experience a non-inertial lift force that pushes them away from the wall \cite{Vlahovska:2009kb}. Reflecting this, most {\it in vitro} studies, as well as numerical \cite{FEDOSOV:2010jm} and theoretical \cite{Olla:1999ck} works, consider interactions between a rigid surface and deformable cells, which adopt an asymmetric shape under flow. Such an asymmetry of the flowing objects is pinpointed as the origin of the lift force arising at the low Reynolds numbers typically encountered in microcirculation. {\it In vivo}, however, blood flow takes place in compliant vessels. In particular, the endothelium (the luminal side of blood vessels) is lined by a glycocalyx, a thin (100-1000 nm) and soft (elastic modulus of 10-100 Pa) layer of polysaccharides bound to the walls and directly exposed to blood flow \cite{Secomb:1998wc}. While the importance of the glycocalyx on blood microrheology is recognized \cite{Damiano:2006ho,Secomb:1998wc}, its quantitative influence on EHD interactions largely remains to be established. More generally, the question of how a thin deformable layer can contribute to ``soft lubrication'' and induce lift forces has been addressed theoretically \cite{Beaucourt:2007ke,Skotheim:2005fy,Urzay:2007gp}, but has received limited attention from the experimental standpoint, with a single study investigating at the macroscopic scale how EHD affects the sliding dynamics of cylinders near a soft wall \cite{Saintyves:2016ic}. In this Letter, we report the investigation of the lift experienced by rigid spherical particles flowing in the vicinity of a surface bearing a polymer brush that mimics the glycocalyx. Using microfluidics and three-dimensional (3D) tracking, we provide the first direct evidence that, under conditions of flow strengths and object sizes relevant to blood circulation, a thin deformable polymer brush gives rise to a sizeable lift on circulating beads, which can be quantitatively described by soft lubrication theory. 

\begin{figure}
\includegraphics[width=\columnwidth]{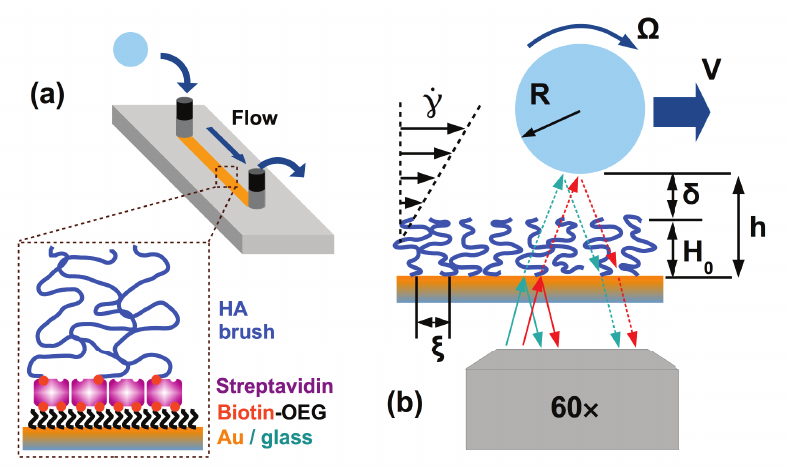}
\caption{\label{Fig-exp} (a) Flow chamber with bottom surface functionalized with a HA brush via biotin/streptavidin binding. (b) A bead traveling in a shear flow of velocity gradient $\dot{\gamma}$. Dual color RICM is used to monitor its distance $h$ from the substrate and its translation velocity $V$.}
\end{figure}

Experiments were performed at room temperature using a parallel-plate flow chamber (Glycotech, USA) composed of a spacer defining a straight channel (Fig. \ref{Fig-exp}a) of rectangular cross-section (height $H=0.250$ mm, width $W=2.5$ mm, length $L=20$ mm), sandwiched between an upper deck with fluid inlet/outlet and a bottom surface consisting of a glass coverslip functionalized with a brush of hyaluronan (HA, the major component of the glycocalyx, see Fig. \ref{Fig-exp}a and details below). The inlet reservoir contained spherical polystyrene beads of radius $R=12.5\, \mu$m (Kisker Biotech, Germany)  suspended in aqueous buffer (10 mM HEPES, pH 7.4, 150 mM NaCl, 2mM CaCl$_2$, viscosity $\eta\simeq10^{-3}$ Pa.s and density $\rho\simeq1000$ kg.m$^{-3}$), while the outlet was connected to a syringe pump (KDS Legato 110) imposing flow rates in the range $Q=1-200 \,\mu$L.min$^{-1}$. Beads were pumped into the channel and left to sediment under quiescent conditions onto the bottom surface of the chamber, after which their motion under imposed flow rate was monitored optically. 
\begin{figure}
\includegraphics[width=\columnwidth]{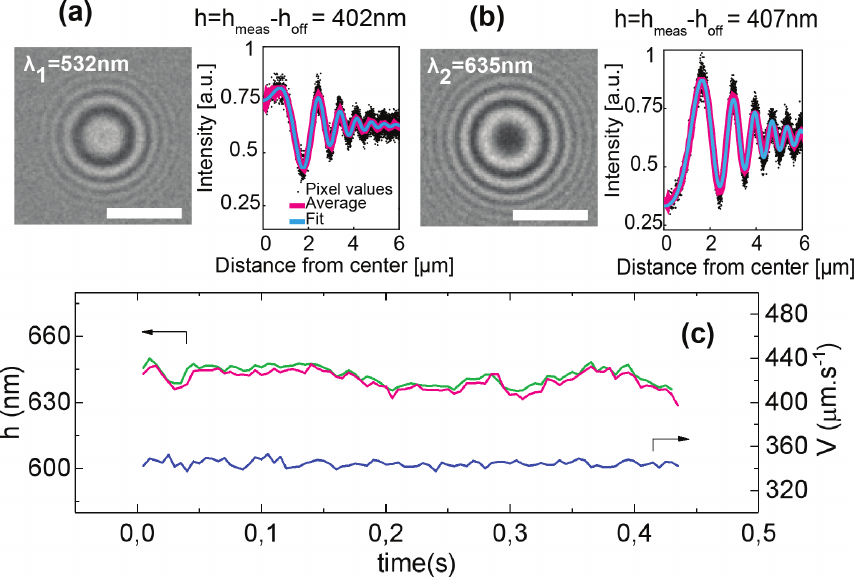}
\caption{\label{Fig-RICM} (a) Left: interference patterns at $\lambda_1=532$ nm (scale bar $5\, \mu$m). Right: radial intensity profile (black dots) extracted from image, azimuthally averaged (magenta line), and fitted with an optical model (cyan line) to determine $h_{\text{meas}}$, from which we compute $h=h_{\text{meas}}-h_{\text{off}}$, with $h_{\text{off}}$ the offset due to the contribution of the gold and streptavidin layers (measured independently, see Supplemental Material). (b) Same as (a) at $\lambda_2=635$ nm. (c) Time series for $h$ (green: $\lambda_1$, magenta: $\lambda_2$) and $V$ (blue), for a bead flowing close to the HA brush.}
\end{figure}
3D tracking was performed by reflection interference contrast microscopy (RICM) using a setup allowing for simultaneous imaging at two wavelengths ($\lambda_1=532$ nm, $\lambda_2=635$ nm,  see Supplemental Material for details). Under flow, the fringe patterns due to interference between the light reflected from the substrate and the surface of the beads were recorded (Fig. \ref{Fig-RICM}a and b) on a camera (ORCA-Flash4.0 Hamamatsu) at rates of up to 200 frames per second. Bead trajectories were analyzed offline, using home-written Labview routines, in order to compute for each $Q$: (i) the steady-state vertical distance $h$ between the substrate and the beads, and (ii) the beads' translation velocity $V$ (Fig. \ref{Fig-exp}b and \ref{Fig-RICM}c). The absolute value of $h$ was determined unambiguously up to $\sim 1.2\, \mu$m owing to the two-color RICM scheme used \cite{Schilling:2004hr}, with an accuracy of $\sim 10$ nm. The in-plane displacements of the beads, from which $V$ was computed, were determined by image correlation with an accuracy of $\sim$50 nm. 

The surface of the coverslip exposed to the flow was functionalized with a layer of HA, as described in \cite{Attili:2012ct,Migliorini:2014ig}. A Ti/Au layer (respectively 0.5 and 5 nm in thickness) was evaporated onto the glass surface. A monolayer of end-biotinylated  oligo(ethyleneglycol) thiols (bOEG-SH) was grafted onto the gold film. A dense layer of streptavidin was bound to the exposed biotin moieties, and further functionalized by incubation with end-biotinylated HA (Fig. \ref{Fig-exp}a). Such a procedure yields HA films stably bound to the substrate in a polymer brush conformation \cite{Attili:2012ct}. To investigate the role of brush thickness and softness, we have studied three different HA surfaces made of chains of well-defined molecular weight (Hyalose, USA):  two substrates obtained by incubating chains of 840$\pm60$ kDa for two different times, yielding high (HA840-h) and low (HA840-l) grafting density samples, and a third one (HA58) bearing a brush made of chains of 58$\pm3$ kDa. Without flow, we measure equilibrium bead heights of respectively $h=405,\, 285$ and $110\pm5$ nm  on the HA840-h, HA840-l and HA58 layers. The gravitational force exerted by a bead sedimented on a brush reads:
\begin{equation}
F_{\text{g}}=\frac{4\pi R^3}{3} g \Delta \rho
\label{gravity}
\end{equation}
With $g=9.81$ m.s$^{-2}$ and a density difference of $\Delta \rho=40$ kg.m$^{-3}$ between the beads and the fluid \cite{Orlishausen:2017gr}, we compute $F_\text{g}=3.2$ pN. This corresponds to an interaction energy per unit area $F_\text{g}/R\simeq2.5\times 10^{-7}$ N.m$^{-1}$ at which we anticipate the brushes to be essentially uncompressed \cite{Attili:2012ct}. Compared to $F_\text{g}$, van der Waals forces between the beads and the glass substrate are negligible at distances $h>10$ nm \cite{Albersdorfer:1999gy}, and repulsive forces of electrostatic origin are screened at the ionic strength used (Debye length $<1$ nm). Therefore, as done previously with similar systems \cite{Albersdorfer:1999gy}, we neglect surface forces and assume that quiescent beads sit at an equilibrium distance from the coverslip that reflects the unperturbed brush height $H_0$. From $H_0$,  we compute (see Supplemental Material) the average distance between the tethered ends of the HA chains (Fig. \ref{Fig-exp}b), $\xi=74\pm17$, $130\pm30$ and $10\pm3$ nm respectively for the HA840-h, HA840-l and HA58 brushes. 
As a control surface, a plain gold-coated coverslip was used, passivated by a layer of bovine serum albumin to minimize non-specific adhesive bead/surface interactions. 
\begin{figure}[htbp]
\includegraphics[width=\columnwidth]{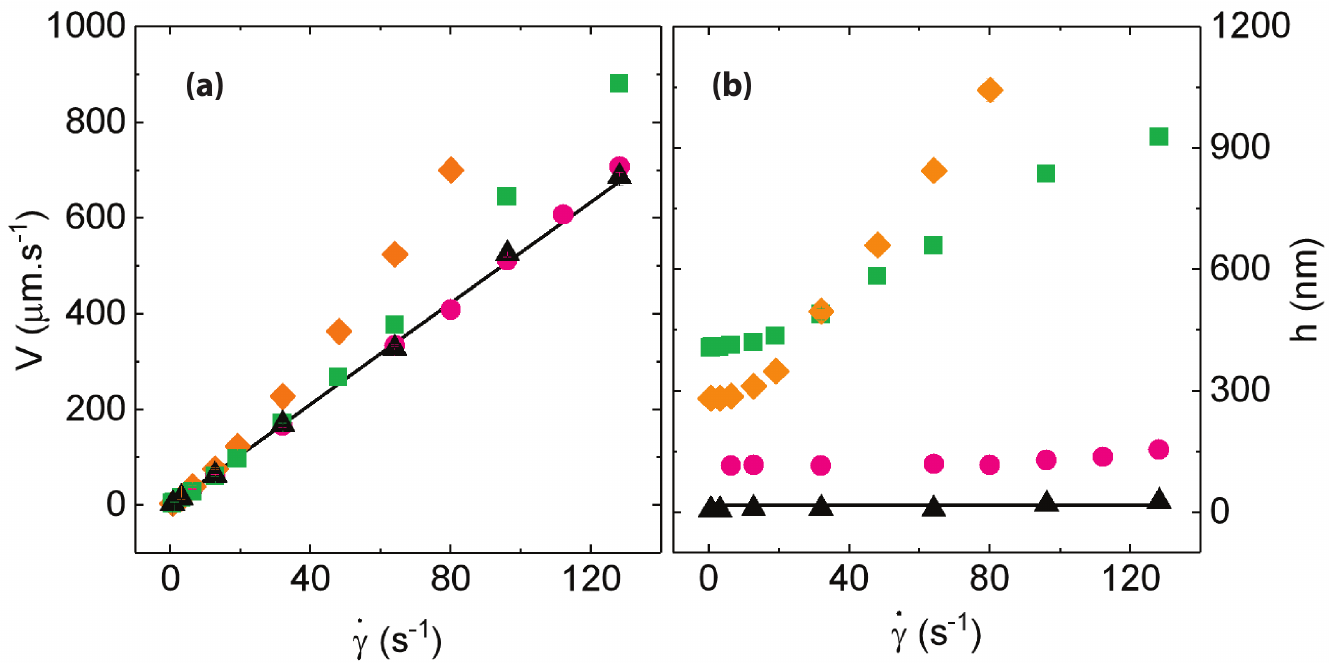}
\caption{\label{Fig-Results} (a)  $V(\dot{\gamma})$ measured on control surface (\textcolor{black}{$\blacktriangle$}),  HA840-h (\textcolor{ForestGreen}{$\blacksquare$}), HA840-l (\textcolor{orange}{$\blacklozenge$}), and HA58 (\textcolor{magenta}{\Large $\bullet$}) brushes. The solid line is the GCB prediction for non-deformable surfaces separated by $20$ nm. (b) Experimentally measured $h(\dot{\gamma})$ (symbols as in (a)). Solid line indicates the constant value of $z=20$ nm used in GCB theory. Error bars accounting for standard error and uncertainty on $h$ and $V$ are about the size of the symbols.}
\end{figure}

An example  time series for $h$ and $V$ of a single bead traveling across the field of view is shown in Fig. \ref{Fig-RICM}c. Single bead data were time-averaged, and measurements over 20 to 50 beads were performed under identical flow conditions to obtain the ensemble-averaged values of $V$ and $h$ shown respectively in Fig. \ref{Fig-Results}a and \ref{Fig-Results}b as a function of wall shear rate $\dot{\gamma}=6Q/(WH^2)$. 

\begin{figure*}[htbp]
\includegraphics[width=18cm]{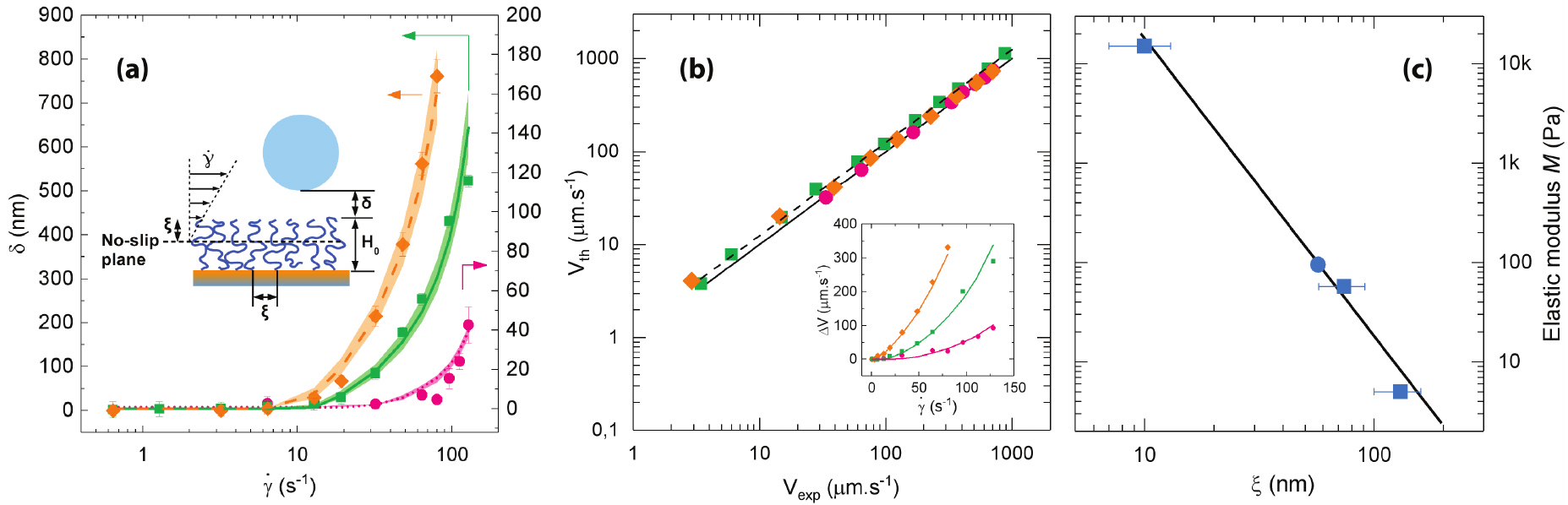}
\caption{\label{Fig-Model} (a) $h(\dot{\gamma})-H_0$ data on HA brushes (symbols as in Fig. \ref{Fig-Results}, vertical scale according to arrows), and theoretical predictions for $\delta(\dot{\gamma})$ with $M=5$ Pa (dashed line), 57 Pa (solid line), and 15000 Pa (dotted line). The shaded area around the theoretical curves is defined by the predictions obtained when $\xi$ is varied from lower to upper bound for each brush. Inset: sketch showing the location of the no-slip plane at $\xi$ below the brush surface. (b) Theoretical ($V_\text{th}$) {\it vs} experimental ($V_\text{exp}$) velocities (symbols as in (a)). The solid line corresponds to  $V_\text{th}=V_\text{exp}$  and the dashed line to $V_\text{th}=1.2V_\text{exp}$. Variations of $V_\text{th}$ due to changes in $\xi$ and error bars on $V_\text{exp}$ are about the symbol size. Inset: measured (symbols) and predicted (lines) deviation from linearity, $\Delta V=V(\dot{\gamma})-S\dot{\gamma}$, with $S$ the slope in the limit of small shear rates. (c) Best fit values  (\textcolor{ProcessBlue}{$\blacksquare$}), measured reference (\textcolor{ProcessBlue}{\Large$\bullet$}) and predictions (solid line) for $M$ as a function of $\xi$. Error bars correspond to the uncertainty on $\xi$.}
\end{figure*}

On the control surface, we observe that $V$ increases linearly with $\dot{\gamma}$, while  $h$  remains small and constant at $15\pm 10$ nm over the range of $\dot{\gamma}$ explored (Fig. \ref{Fig-Results}, triangles). When beads are flowing past the HA840-h brush, their velocity increases linearly with $\dot{\gamma}$ and $h$ remains close to $H_0$ at shear rates below $\sim$20-30 s$^{-1}$. However, for $\dot{\gamma}>$30 s$^{-1}$, $V$ grows more than linearly with $\dot{\gamma}$ while $h$ steadily increases and reaches up to 900 nm at the largest shear rate (Fig. \ref{Fig-Results}, squares). Such a trend is further amplified with the HA840-l brush (Fig. \ref{Fig-Results}, diamonds). With the HA58 brush, we observe that $V$ grows quasi-linearly with $\dot{\gamma}$, while $h$ increases by only $40$ nm above $H_0$ at the largest $\dot{\gamma}$ (Fig. \ref{Fig-Results}b, and magnified in Fig. \ref{Fig-Model}a). Thus, there is a lift of the beads away from the brushes, with a magnitude depending on the shear rate and the type of brush. We now discuss the possible origins of such a phenomenon.

Given the low Reynolds number in our experiments ($Re\lesssim 10^{-2}$), we first compare the results from the control experiment with the theory of Goldman, Cox and Brenner (GCB) for a rigid bead in a shear flow past a non-deformable surface \cite{Goldman:1967ur}. For bead/surface distances $z\ll R$, GCB computed the following bead translation ($V$)  and angular ($\Omega$) velocities (Fig. \ref{Fig-exp}b) \footnote{Numerical coefficients slightly differ from the original formula by GCB. We have fitted the values tabulated in reference \cite{Goldman:1967ur} and found that the present set of coefficients yields a better agreement than those provided in the original publication.}:
\begin{equation}
V=\dot{\gamma}R\frac{(1+z/R)}{0.7625-0.2562\ln(z/R)}
\label{GCB1}
\end{equation}
\begin{equation}
\Omega=\frac{\dot{\gamma}}{1.6167-0.4474\ln(z/R)}
\label{GCB2}
\end{equation}
Using Eq. (\ref{GCB1}), we obtain excellent agreement between GCB theory and our data on the control surface when setting $z=20$ nm  (solid lines in Fig. \ref{Fig-Results}), consistent with the measured $h$. The results of our control experiment therefore match very well the predictions for a rigid sphere flowing in quasi-contact with a rigid plane.

To address the lift from the HA surfaces, we first consider inertial forces. Even at low $Re$, it has been shown that an inertial lift force can act on a bead moving close to a wall in a linear shear flow \cite{Krishnan:1995je,Cherukat:2006ga}. Cherukat and McLaughlin have computed an expression for this inertial lift force, valid in the limit $z\ll R$ \cite{Cherukat:2006ga}:
\begin{equation}
F_{\text{in}}=\rho R^2 V_r^2 I(\Lambda_G,\, \kappa)
\label{inertia}
\end{equation}
where $V_r=V-\dot{\gamma}(R+z)$ is the difference between the bead velocity and the fluid velocity at the location of the bead center of mass, and $I(\Lambda_G,\, \kappa)$ is given by
\begin{align}
I=&[1.7669+0.2885\kappa-0.9025\kappa^2+0.507625\kappa^3] \label{inertia2} \\ 
&-[3.2415/\kappa+2.6729+0.8373\kappa-0.4683\kappa^2]\Lambda_G \notag \\
&+[1.8065+0.89934\kappa-1.961\kappa^2+1.02161\kappa^3]\Lambda_G^2 \notag
\end{align}
with $\Lambda_G=\dot{\gamma}(R+z)/V_r$ and $\kappa=R/(R+z)$. Taking $V=900\, \mu$m.s$^{-1}$, $\kappa \simeq 1$, and $\dot{\gamma}=128$ s$^{-1}$, we estimate the maximum inertial lift force $F_{\text{in}}\simeq 1.9$ pN, which is lower than $F_\text{g}$. Therefore, inertial effects alone cannot induce lift in the range of shear rates explored here, in agreement with our control experiment.

With electrokinetic effects being negligible at the ionic strength used here (see Supplemental Material), the only other mechanism that can lead to lift is due to EHD: in the presence of a thin and soft surface layer, elastic deformations induced by the pressure field in the lubricating fluid are predicted to give rise to a lift force ($F_{\text{EHD}}$) on a rigid sphere \cite{Beaucourt:2007ke,Skotheim:2005fy,Urzay:2007gp}. To test whether this accounts for our observations, we start from the expression derived by Urzay et al. \cite{Urzay:2007gp} for $F_{\text{EHD}}$:
\begin{equation}
F_{\text{EHD}}=\frac{\eta{^2}R{^2}H_0V_s{^2}}{M \delta^3}\left( \frac{48\pi}{125} + \frac{4\pi(19+14\omega)}{25(1+\omega)}\frac{\delta}{R}\right)
\label{EHD}
\end{equation}
with $\eta$ the fluid dynamic viscosity, $H_0$ the layer thickness, $M$ its longitudinal elastic modulus \footnote{$M=E(1-\nu)/[(1+\nu)(1-2\nu)]$, with $E$ the Young modulus and $\nu$ the Poisson ratio.}, $\delta$ the bead/layer distance (see Fig. \ref{Fig-exp}b), $V_s=V-\Omega R$, and $\omega=-\Omega R/V$. 

At a given shear rate, the steady-state value of $\delta$ is set by the balance of vertical forces on a bead:
\begin{equation}
F_{\text{EHD}}+F_{\text{in}}=F_{\text{g}}
\label{balance}
\end{equation}
Eq. (\ref{EHD}) was derived for an impermeable elastic layer with a no-slip boundary condition at its surface, but holds for a poroelastic layer, under conditions on $M$ discussed below \cite{Skotheim:2005fy}. Moreover, a polymer brush in a shear flow is expected to be penetrated by the flow over a distance of order $\xi$, the inter-chain spacing \cite{Harden:1997cz,Lee:2012el,Deng:2012ht}. We account for this by assuming the no-slip plane to lie at a distance $\xi$ below the top of the layer (Fig. \ref{Fig-Model}a inset). In the spirit of previous works on hydrodynamic interactions with finite slip length \cite{Baudry:2001jf,CottinBizonne:2002br,Charrault:2016gq}, we then replace $\delta$ by $\delta+\xi$ and $H_0$ by $H_0-\xi$ (the thickness of the layer not penetrated by the flow) in Eq. (\ref{EHD}). Similarly, we set $z=\delta+\xi$ in Eqs. \ref{GCB1}$-$\ref{inertia2}, and use GCB results to compute $V_r(\dot{\gamma},\, \delta)$, $V_s(\dot{\gamma},\, \delta)$, and $\omega(\dot{\gamma},\, \delta)$ (the above assumptions are further discussed in Supplemental Material).
The theoretical values for the lift  can then be determined as a function of $\dot{\gamma}$ by solving Eq. (\ref{balance}) for $\delta$, knowing $R$, $\rho$, $\Delta\rho$, $\eta$, $\xi$ and $H_0$.

The resulting predictions for $\delta$ are compared with the data for $h-H_0$ in Fig. \ref{Fig-Model}a. Keeping only $M$ as an adjustable parameter, we obtain a very good agreement between the measured and computed lift, with $M=15000$, 57 and 5 Pa respectively for the HA58, HA840-h and HA840-l brushes, with a marginal effect due to the uncertainty on $\xi$. As observed experimentally, the model predicts forces that are too low to induce lift below $\dot{\gamma}\sim 10-30$ s$^{-1}$, while for larger shear rates the lift gradually increases, an effect that is augmented with decreasing brush modulus.
Eliminating the inertial term in the force balance to account only for EHD demonstrates that inertial effects become significant for $\dot{\gamma}\geq 50$ s$^{-1}$, and that EHD alone accounts for about 75\% of the lift observed at the highest shear rate (see Supplemental Material). 
Besides, the velocities predicted at the corresponding $\dot{\gamma}$ and $\delta$ agree to within 20\% or better with the experimental data (Fig. \ref{Fig-Model}b), and the non-linearity of the $V(\dot{\gamma})$ curves is quantitatively captured by the model (Fig. \ref{Fig-Model}b inset). 

Furthermore, the values of $M$ required to reproduce the data are in agreement with previous findings on the mechanical properties of HA brushes. As for polymer gels, the modulus of a brush scales as $M\sim 1/\xi^3$ (see Supplemental Material). From previous measurements of the mechanical response of HA brushes \cite{Attili:2012ct}, we determine $M_{\text{ref}}\simeq 100$ Pa for a brush of $\xi_{\text{ref}}=57$ nm (see Supplemental Material). We can thus compute the expected moduli of the brushes from $M=M_{\text{ref}}/(\xi/\xi_{\text{ref}})^3$, and compare them to the above best fit values. As shown in Fig. \ref{Fig-Model}c, we obtain a good agreement between the moduli, strengthening the fact that EHD does govern the observed behaviors. Now, $M_{\text{ref}}$ comes from quasi-static measurements and corresponds to the ``drained'' value of the modulus, determined under conditions where water is free to flow in the brush and does not contribute to the stiffness. Evaluating the poroelastic time of the brushes, $\tau_\text{p}\sim \eta H_0^2/(M\xi^2)$, and the experimental time scale $\tau_\text{exp}\sim \sqrt{\delta R}/V$ \cite{Skotheim:2005fy}, we find that $\tau_\text{exp}\gg \tau_\text{p}$, irrespective of the brush or flow conditions. Following the argument given in  \cite{Skotheim:2005fy}, this confirms that the drained moduli should indeed be used in Eq. (\ref{EHD}).

In summary, our work shows how a compliant biomimetic layer affects the near-wall motion of microparticles. Our observations are quantitatively supported by theoretical predictions based on EHD, thus providing direct evidence of soft lubrication at play at small scales. This is likely to have significant influence on the behavior of red blood cells in blood circulation. Indeed, a RBC ($R\simeq3\, \mu$m) flowing in plasma ($\eta\simeq 1.5$ mPa.s) under a physiological shear rate $\dot{\gamma}\simeq100$~s$^{-1}$ , at a distance $\delta\simeq 0.5\,\mu$m from a $\mu$m-thick glycocalyx, would experience a force $F_\text{EHD}\simeq 0.15$~pN due to glycocalyx softness (which dominates over that of adjacent tissues, see Supplemental Material). From a recent study of the drift velocity $v_z$ of RBCs under shear \cite{Grandchamp:2013jq}, we compute  $v_z=\beta\dot{\gamma}/(R+\delta)^2\simeq 3\, \mu$m.s$^{-1}$ at the same $\delta$ and $\dot{\gamma}$, with $\beta\simeq0.36\, \mu$m$^3$ determined experimentally \cite{Grandchamp:2013jq}. This translates into a lift force due to cell deformation $F_\text{cell}\sim 6\pi\eta R v_z\simeq 0.25$~pN. It thus appears that the contributions of cell and wall deformations to lubrication forces are of comparable magnitude at sub-$\mu$m distances from the wall. In conclusion, the present study underlines the important, yet often overlooked, mechanical role that the soft endothelial glycocalyx is likely to play in regulating cell/wall interactions in blood flow.

\begin{acknowledgments}
We acknowledge the ``Emergence'' and ``AGIR'' programs of Universit{\'e} Grenoble Alpes, the Spanish Ministry for Economy and Competiveness (project MAT2014-54867-R, to R.P.R.), the European Research Council (Starting grant 306435, to R.P.R.), the French Agence Nationale de la Recherche (Grant ANR-13-JS08-0002-01, to L.B.), and the Centre National d'Etudes Spatiales (CNES) for funding. We acknowledge the ``Prestige'' European program and the CNES for fellowships (H.D.). We are grateful to Liliane Coche-Guerente (Department of Molecular Chemistry, Grenoble) for assistance with surface functionalization/characterization, to Luis Yate (CIC biomaGUNE) for gold coatings, and to Gwennou Coupier (LIPhy, Grenoble) for stimulating discussions regarding the lift of RBCs. 
\end{acknowledgments}


%

\newpage

\onecolumngrid

\newcommand{\snum}{S}

\renewcommand{\thefigure}{S\arabic{figure}}
\setcounter{figure}{0}

\renewcommand{\theequation}{\snum \arabic{equation}}
\setcounter{equation}{0}

\renewcommand{\thepage}{S\arabic{page}}
\setcounter{page}{1}

{\Large \bf \begin{center} Supplemental Material \end{center}}

\section{RICM setup and data analysis}

RICM images were acquired on a modified inverted microscope (IX71 Olympus, Japan). Custom-built illumination incorporated an incoherent white-light source (HPLS345, Thorlabs), a green/red dual-band interference filter (FF01-534/635-25, Semrock, USA) and a calibrated aperture diaphragm (Thorlabs). Illumination light was reflected on a broadband polarizing beamsplitter cube (Thorlabs), passed through a precision broadband quarter-waveplate (Fichou, France) and was focused onto the sample using a super-apochromatic objective (UPLSApo 60XO, Olympus, Japan). Reflected light was collected through the same objective, passed a second time through the quarter-waveplate and was then transmitted through the polarizing cube, allowing separation from the incoming light. The RICM signal was then imaged on a sCMOS camera (ORCA-Flash 4.0 V2, Hamamatsu, Japan) using relay lenses (ITL200, Thorlabs) and dichroic beamsplitters and filters to separate the green and red images (FF560-FDi01-25x36, FF01-531/46-25 and  FF01-629/56-25, Semrock, USA). In order to accommodate for fast displacement of the beads at high shear rates, both images were acquired simultaneously on different regions of the camera.

Analysis of the RICM pattern was performed as follows: the radially symmetrical pattern was azimuthally averaged to obtain an intensity curve as a function of the distance $r$ to the bead lowest point. This curve can then be fitted with the following function:
\begin{equation}
I(r)=A_1\exp^{-\frac{r^2}{w_1^2}}+A_2\exp^{-\frac{r^2}{w_2^2}}\cos(\frac{4\pi n_{\mathrm{buffer}}}{\lambda}(h_{\mathrm{meas}}+\delta h(r)))
\label{fitRICM}
\end{equation}
where $A_1$, $w_1$, $A_2$ and $w_2$ account for the amplitude of the signal and the fringes, and their empirical decrease due to spatial and temporal decoherence and reflection on the top surface of the bead \cite{Raedler:1992he}. $n_\text{buffer}$ is the refractive index of the liquid medium, and $\delta h(r)$ accounts for the change in optical distance between the bead and the coverslip as a function of the radial coordinate, with $\delta h(r=0)=0$. 

Finally, $h_{\mathrm{meas}}$ incorporates both the optical distance between the lowest point of the bead and the top of the gold layer, and the phase shifts of the effective reflection and transmission coefficients of the gold layer resulting from multiple reflections at its boundaries, affecting both the effective reflection of the gold layer and that of the bead (multiple reflections at the surface of the bead are negligible due to the small value of the reflection coefficient ($<1\%$)). 
Being highly sensitive to the exact gold layer thickness, such shifts are difficult to evaluate precisely. However, because the different phase terms add linearly, the resulting phase offset can be determined experimentally from the difference between $h_{\mathrm{meas}}$ values measured in the presence and in the absence of the HA brush. To this aim, beads were added to a surface functionalized with the streptavidin monolayer prior to the addition of end-biotinylated HA, and RICM patterns were analyzed for $\sim 20$ beads to obtain a reference value $h_{\mathrm{off}}$. This reference was subsequently subtracted from $h_{\mathrm{meas}}$ obtained in the presence of the HA brush to compute the height of the bead with respect to the bottom of the HA brush. The optical contribution of the brush was neglected in this calculation due to its very low density, corresponding to refractive index changes of $\approx2\times 10^{-4}$. 

From equation~(\ref{fitRICM}), it appears that the RICM pattern is identical for all heights $h_{\mathrm{meas}}$ separated by $\lambda/(2n_{\mathrm{buffer}})\approx200$ nm. In order to determine the correct height of the bead, two illumination wavelengths $\lambda_1$ and $\lambda_2$ are used, thereby limiting the ambiguity to height values separated by $\lambda_1 \lambda_2/(2 n_{\mathrm{buffer}}(\lambda_2-\lambda_1))$. Here $\lambda_1=532$ nm and $\lambda_2=635$ nm, and hence $\lambda_1 \lambda_2/(2 n_{\mathrm{buffer}}(\lambda_2-\lambda_1))\approx 1.2 \mu m$ \cite{Schilling:2004hr}. Below this height, dual-colour RICM patterns can be analyzed unambiguously. 

Only beads separated by about 10 radii or more from their closest neighbors were analyzed in our study, in order to avoid hydrodynamic coupling between beads and collective effects.

\section{Estimation of inter-chain average distance $\xi$}

In a previous study \cite{Attili:2012ct}, we have shown that force/distance curves measured by colloidal-probe AFM on HA brushes at sufficiently high ionic strength could be quantitatively described by self consistent field theory (SCFT) of neutral polymer brushes (by treating the screened electrostatic interactions as an enhanced effective excluded volume), from which the molecular parameters of the brushes could be retrieved. We use the same framework here in order to deduce $\xi$ from the brush height $H_0$. According to SCFT, the relationship between these two quantities reads \cite{Attili:2012ct}:

\begin{equation}
H_0=\left( \frac{8}{\pi^2}\right)^{1/3}\left( \frac{p\nu}{b\xi^2}\right)^{1/3} l_c
\label{SCFT}
\end{equation}
where $l_c$ is the contour length of the chains, $b= 1.0$ nm is the size of the monomer, $\nu$ is the excluded volume, and $p$ is a non-dimensional chain stiffness parameter. 

The contour length of the chains are deduced from their molecular weight $M_w$, with $M_0=378$ g.mol$^{-1}$ the monomer molar mass: $l_c=bM_w/M_0$. This yields $l_c\simeq2220\pm160$ nm for HA840, and $l_c\simeq155\pm8$ nm for the HA58 sample. Using the previous determination of $(p\nu/b)^{1/3}\simeq3.45\pm0.2$ nm$^{2/3}$ \cite{Attili:2012ct}, we then compute $\xi$ from $H_0$ with Eq. (\ref{SCFT}) and obtain the values reported in the main text, with the uncertainty on $\xi$ associated with the error bars on the above parameters.

\section{Magnitude of electroviscous effects}

It is known that electrokinetic effects may give rise to repulsive interaction forces under flow when two electrically charged surfaces are moving past each other in an electrolyte solution \cite{Bike:1995hc}. In order to ensure that such a phenomenon is negligible in our study, we use a recently developed theory to compute the expected electroviscous lift forces under our experimental conditions. Tabatabaei et al. \cite{Tabatabaei:2006hk} have derived the following expression for the electroviscous lift force $F_{ev}$ acting between a charged planar surface and a charged bead translating and rotating at a small distance from the wall :

\begin{equation}
F_\text{ev}=\frac{48\pi}{25} R^2 c_{\infty}kT \varepsilon^4 \text{Pe}^2 \left(\frac{\delta}{R} \right)^{-2}\left[ (G_p+SH_p+G_w+SH_w)^2(1-\omega)^2 -\alpha[(G_p+SH_p)^2-(G_w+SH_w)^2](1-\omega^2) \right]
\label{electro}
\end{equation}
with $R$ the bead radius, $\delta$ the bead/surface distance, $c_\infty$ the bulk number density of ions, $kT$ the thermal energy, $\varepsilon$ the ratio of the Debye length to the bead radius, $S=D_1/D_2$ the ratio of the diffusion coefficients of the counterions ($D_1$, positive species for a negatively charged surface) and the co-ions  ($D_2$), and the coefficient $\alpha\approx -1.667$. The Peclet number is defined as $\text{Pe}=RV/D_1$, with $V$ the bead translational velocity. As in the main text, $\omega=-R\Omega/V$, with $\Omega$ the rotation speed. The quantities $G_i$ and $H_i$ are defined as:

\begin{equation}
G_i=\ln\frac{1+e^{-\tilde{\psi}_i/2}}{2}, \,\,\, H_i=\ln\frac{1+e^{\tilde{\psi}_i/2}}{2}
\label{electro2}
\end{equation}
where $i=(w, p)$ stands for wall and particle. The reduced surface potentials $\tilde{\psi}_i$ are such that $\psi_i=\tilde{\psi}_ikT/(ze)$, with $e$ the elementary charge, $z$ the ion valence, and $\psi_i$ the surface potentials.

We take $D_1=1.33\times 10^{-9}$ m$^2$.s$^{-1}$ and $D_2=2\times 10^{-9}$ m$^2$.s$^{-1}$ for the Na$^+$ and Cl$^-$ ions, respectively \cite{Nielsen:1952dn}.

From the effective surface charge density of HA brushes computed in reference \cite{Attili:2012ct}, we can estimate the polymer layers studied here to carry a negative charge density in the range $\sigma=6\times 10^{-5}-8\times 10^{-4}$ C.m$^{-2}$. At a NaCl concentration of 150 mM, hence a Debye length of $\lambda_D=0.8$ nm, this translates into a brush surface potential of $\psi_w\simeq\sigma \lambda_D/(\epsilon_r\epsilon_0)$ between -0.07 and -1 mV. Besides, we assume a typical value of $\psi_p=-50$ mV for the polystyrene beads.

\begin{figure*}
\includegraphics[width=16cm]{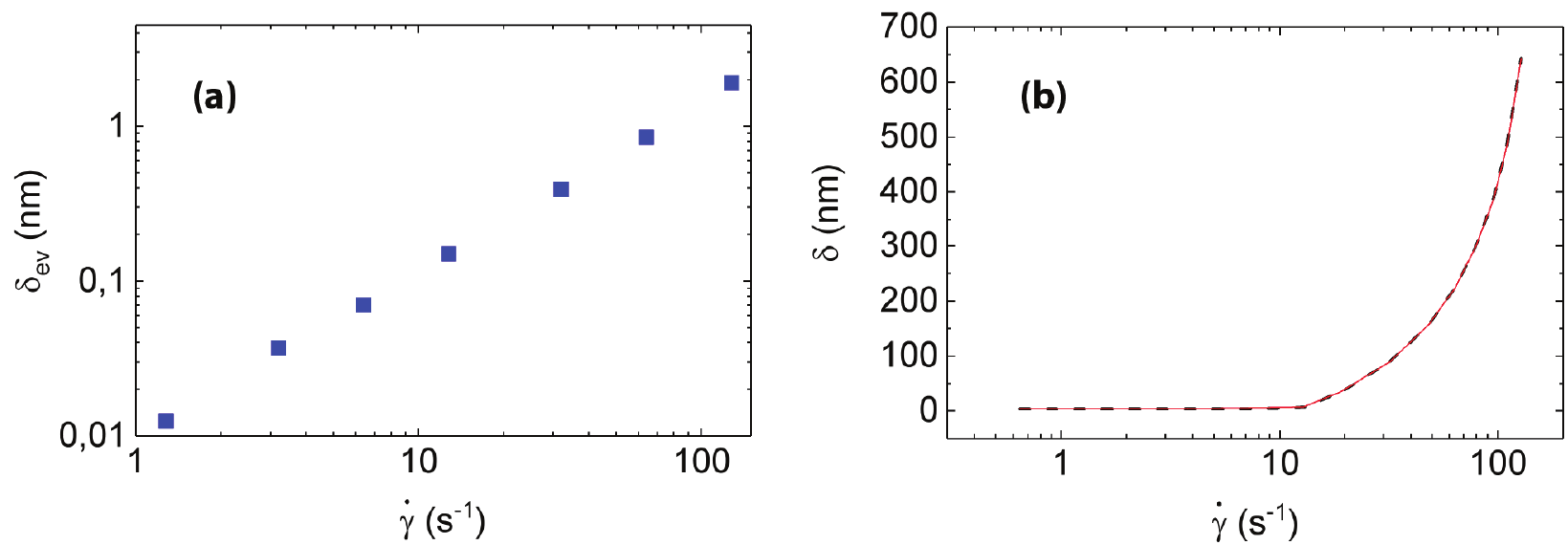}
\caption{\label{Fig-EV} (a) Predicted lift due to electrokinetic effects only ($\delta_\text{ev}$), as a function of $\dot{\gamma}$. Results obtained for $\psi_w=-0.07$ and -1 mV cannot be distinguished. (b) Comparison of the lift predicted by accounting (dashed line) or not (red solid line) for electrokinetic effects in the total force balance.}
\end{figure*}

Using the expressions derived by Goldmann, Cox and Brenner to compute $V$ and $\Omega$ as a function of $\delta$ and the imposed shear rate (see main text) \cite{Goldman:1967ur}, we write the vertical force balance on a bead, $F_\text{ev}=F_\text{g}$ (with $F_\text{g}$ the gravity force), and solve for $\delta$ in order to estimate the magnitude of the electroviscous effects alone. As can be seen on Fig. \ref{Fig-EV}a, the electrokinetically-induced lift of a bead is, at the high ionic strength used here, typically two orders of magnitude below what we observe experimentally. This is further demonstrated on Fig. \ref{Fig-EV}b, where it can be seen that the total lift predicted from a force balance incorporating all possible lift mechanisms is indistiguishable from that computed by ignoring $F_\text{ev}$ in the force balance.

\section{Modeling assumptions and brush deformation}

As mentioned in the main text, the HA brushes placed in a shear flow are expected to be penetrated by the flow over a thickness of order $\xi$. A full EHD modeling of our experimental results would therefore require a theory describing the motion of a bead driven by a linear shear flow in the vicinity of a poroelastic layer. In the absence of such a theory, we rely on a set of simplifying assumptions in order to adapt the existing EHD framework to the situation of interest here:

(i) We consider that the expression derived by Urzay et al. \cite{Urzay:2007gp} for the lift force $F_\text{EHD}$, reproduced below, holds for a poroelastic layer, as discussed qualitatively in references \cite{Skotheim:2005fy,Gopinath:2011fd}.

\begin{equation}
F_{\text{EHD}}=\frac{\eta{^2}R{^2}H_0 V_s{^2}}{M \delta^3}\left( \frac{48\pi}{125} + \frac{4\pi(19+14\omega)}{25(1+\omega)}\frac{\delta}{R}\right)
\label{EHD}
\end{equation}

(ii) We assume that the layer thickness is $H_0-\xi$. This implies, as done in previous studies \cite{Stuart:1984dz}, that the hydrodynamic thickness of the polymer layer is the relevant one.

(iii) Consistently with (ii), we assume that the relevant bead/surface distance in $F_\text{EHD}$ is $\delta+\xi$. We thus treat the penetration length $\xi$ as an effective slip length, as done in a previous theoretical analysis of squeeze flow with polymer brushes \cite{Potanin:1995kz}, and make the simple hypothesis that the lubrication force between the surfaces in the presence of slip can be computed from the no-slip case by merely shifting the sphere/wall distance by $\xi$. Such an approximation has been used in several earlier works \cite{Lee:2002cj,Baudry:2001jf,CottinBizonne:2002br}, and has been found to hold even for small sphere/wall distances \cite{CottinBizonne:2002br}.

(iv) We rely on Goldman-Cox-Brenner (GCB) theory in order to compute the bead translation ($V$) and rotation ($\Omega$) velocities as a function of the imposed shear rate $\dot{\gamma}$. As in (iii), we further assume that the relevant bead/wall distance is $\delta + \xi$. Using numerical results obtained by Damiano et al. \cite{Damiano:2004iw}, who computed $V$ and $\Omega$ for a bead flowing past a porous layer of permeability $\xi^2$, we have checked that our assumption is indeed consistent with their results.\\
Assumption (iv) accounts only for a finite slip length, but does not take into consideration any effect of the layer deformability on velocities. However, based on a recent theoretical work addressing the role of an elastic layer on the drag and rotation of a sliding cylinder \cite{Rallabandi:2017id}, we expect corrections to $V$ and $\Omega$ due to elasticity to be of order $\Lambda^2\simeq 5\%$ (with $\Lambda\simeq\eta \dot{\gamma} H_0 R^{3/2}/(M\delta^{5/2})$), {\it i.e.} to remain small compared to the case of rigid bodies treated by GCB.

\vspace{5mm}

This set of assumptions allows us to capture quantitatively the observed variations of  $\delta$ with  $\dot{\gamma}$. The model is robust and only weakly sensitive to the exact value of $\xi$, which can be varied between its upper and lower bounds without compromising the agreement between experimental data and predictions. 

\begin{figure*}
\includegraphics[width=8cm]{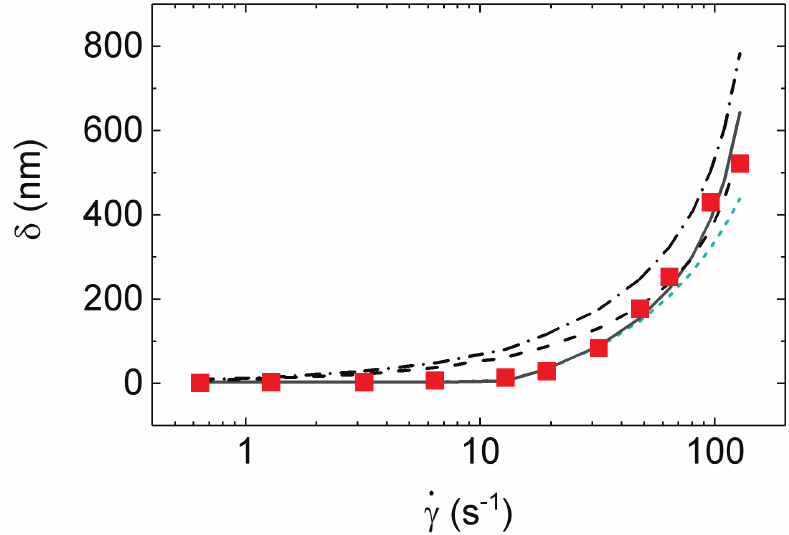}
\caption{\label{Fig-testmodelSI} $\delta$ $vs$ $\dot{\gamma}$ measured on HA840-h brush (\textcolor{red}{$\blacksquare$}), predicted with $M=57$ Pa and $\xi=74$ nm (solid line), $M=57$ Pa and $\xi=0$ (dash-dotted line), $M=120$ Pa and $\xi=0$ (dashed line), $M=57$ Pa and $\xi=74$ nm without inertial force (blue  dotted line). }
\end{figure*}

Furthermore, as mentioned in the main text, we show in Fig. \ref{Fig-testmodelSI} that the inertial term retained in the force balance becomes significant at shear rates above $\sim50$ s$^{-1}$ only, and that EHD alone accounts for most of the observed lift even at the largest shear rates.

For the sake of completeness, we also illustrate in Fig. \ref{Fig-testmodelSI} the effect of setting the slip length to zero in the model ($i.e.$ assuming an elastic and impermeable layer with a no-slip condition at its surface).
It can be seen that, keeping the elastic modulus constant, the predicted values of $\delta$ lie, as expected, above those obtained with a slip length of $\xi$. We can however predict the correct trend and magnitude for the lift at large shear rates by using a modulus approximately twice as large as the one needed with finite slip. Still, we observe that  the no-slip assumption does not capture the low shear rate regime, where the model overestimates the lift. Overall, we expect the no-slip case and the simple slip-length model to represent respectively an upper and lower limit of the predictions that a full poroelastic treatment would yield.

\begin{figure*}
\includegraphics[width=8cm]{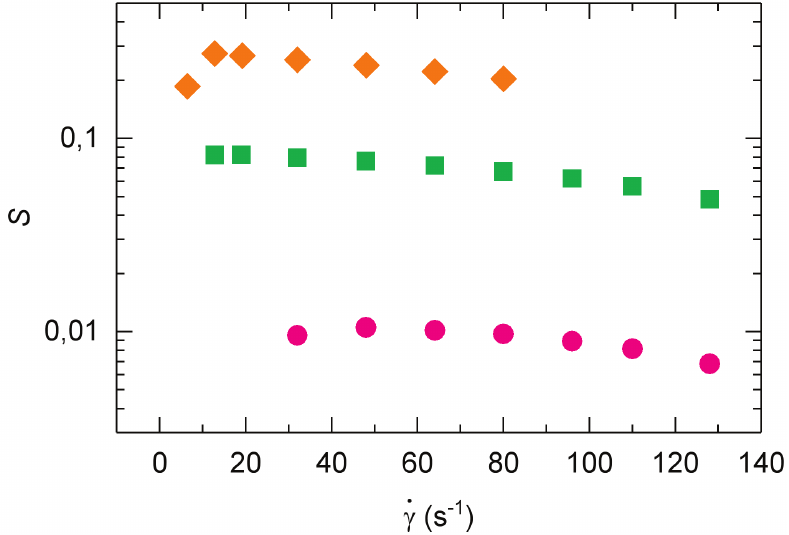}
\caption{\label{Fig-strainSI} Brush strain $\mathcal{S}$ $vs$ $\dot{\gamma}$ computed for the HA840-h (\textcolor{ForestGreen}{$\blacksquare$}), HA840-l (\textcolor{orange}{$\blacklozenge$}), and HA58 (\textcolor{magenta}{\Large $\bullet$}) brushes.}
\end{figure*}

Finally, we estimate the maximum brush strain, $\mathcal{S}$, induced by EHD. Following reference \cite{Urzay:2007gp}, this is given by:

\begin{equation}
\mathcal{S}=\frac{\eta R^{1/2}V(1+\omega)}{M \delta^{3/2}}
\label{strain-EHD}
\end{equation}
It can be seen on Fig. \ref{Fig-strainSI}, where we plot $\mathcal{S}(\dot{\gamma})$, that the typical strain of the brush due to EHD is below 1, 10 and 30\% for the HA58, HA840-h and HA840-l respectively. As shown in the next section, the mechanical response of HA brushes can be described with a constant elastic modulus up to strains of $\sim30\%$, suggesting that non-linear elasticity can be neglected for the three samples studied, including the softest one.

\section{Estimation of the  longitudinal elastic moduli of the brushes}

\begin{figure*}[h!]
\includegraphics[width=12cm]{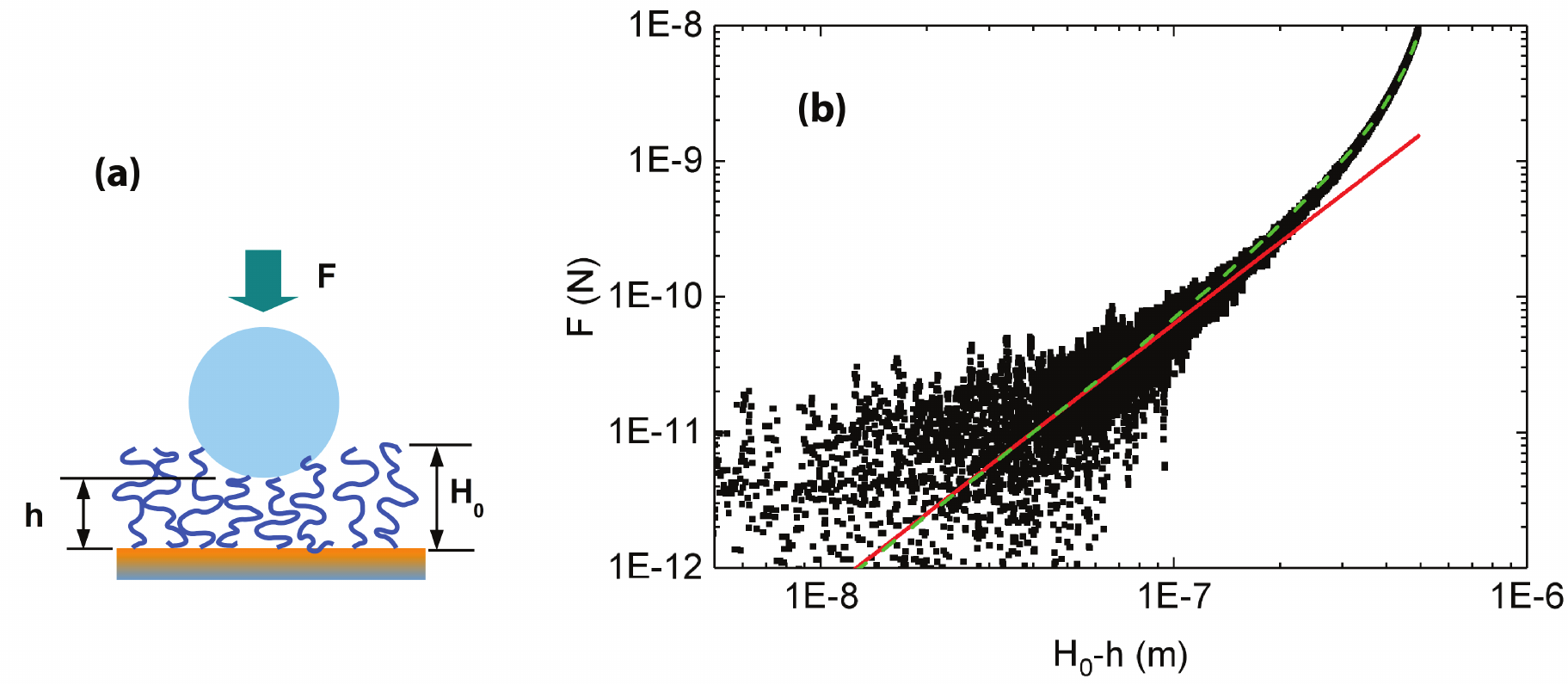}
\caption{\label{Fig-Hertz} (a) Sketch of brush of unperturbed thickness $H_0$ compressed to a thickness $h$ upon application of a force $F$ on the bead. (b) Experimental data $F(H_0-h)$ ($\blacksquare$) and fit of the quadratic regime (red solid line) using Eq. (\ref{hertz}), and full fit using Eq. (\ref{degennes}) with $H_0=590$ nm and $\xi=57$ nm (green dashed line).}
\end{figure*}

We first come back on the analysis of the force/distance curves ($F(h)$, see Fig. \ref{Fig-Hertz}a) of HA brushes reported  in reference  \cite{Attili:2012ct}. We note that the contact mechanics situation involved in the colloidal-probe AFM-RICM experiments reported in  \cite{Attili:2012ct} corresponds to that described by Johnson as the ``elastic foundation model'', for which the relationship between force ($F$) and indentation of the layer ($H_0-h$) is predicted to be \cite{Johnson:1985fo}:

\begin{equation}
F=\frac{\pi M (H_0-h)^2 R}{H_0}
\label{hertz}
\end{equation}
with $R$ the bead radius and $M$ the longitudinal modulus of the elastic layer \cite{Gacoin:2006hh}. 

In Fig. \ref{Fig-Hertz}b, we have plotted $F(H_0-h)$ from a dataset taken from \cite{Attili:2012ct}, measured on a HA brush of $\xi_\text{ref}=57$ nm and $H_0=590$ nm immersed in a 150mM NaCl solution. It can be seen on the double-logarithmic scale of the figure that, for $(H_0-h)\lesssim 200$ nm, the experimental curve exhibits a regime where $F$ increases as $(H_0-h)^2$. Fitting this quadratic regime with Eq. (\ref{hertz}) yields $M_\text{ref}\simeq 100$ Pa, in excellent agreement with the outcome of a slightly different analysis proposed previously for these HA brushes \cite{Attili:2013kz}. 

We now derive the expected scaling for $M$ as a function of $\xi$. Within the Alexander-deGennes model of polymer brushes, which accurately describes the mechanical response of our systems (Fig. \ref{Fig-Hertz}b), we expect $F(h)$ to be given by \cite{Attili:2012ct}:

\begin{equation}
F\simeq R\frac{H_0 kT}{\xi^3}\left[7\left(\frac{H_0}{h}\right)^{5/4}+5\left(\frac{h}{H_0}\right)^{7/4}-12\right]
\label{degennes}
\end{equation}

Equating expressions (\ref{hertz}) and (\ref{degennes}), we obtain for the elastic modulus, in the limit of small deformations $h/H_0\rightarrow 1$, and dropping numerical prefactors:

\begin{equation}
M\sim \frac{kT}{\xi^3}
\label{scalingM}
\end{equation}
which is the same scaling obeyed by polymer gels \cite{Rubinstein:2003vd}.

\section{Effect of substrate elastic properties}

In order to determine to what extent the mechanical properties of the substrate underlying the HA brush affect the EHD interactions, we use the criterion derived by Leroy and Charlaix in their theoretical analysis of non-contact probing of thin elastic films \cite{LEROY:2011ck}. They have shown that, for a film of thickness $H_0$ and elastic modulus $E$ placed on a substrate of modulus $E_s$, the substrate contributes to less than 1\% of the lubrication forces provided that:

\begin{equation}
H_0\gtrsim 2 \left( \frac{E}{E_s}\right)^{1/3}\sqrt{R\delta}
\label{criterion}
\end{equation}

Taking $R=12.5\, \mu$m and $E_s\simeq 60$ GPa for the glass substrate, we compute the right-hand-side of Eq. (\ref{criterion}) to be at most 2--10 nm for the three HA brushes, showing that the contribution of the glass substrate is negligible in our experiments.

If we now estimate the same criterion for a blood cell of $R\sim 3\, \mu$m flowing at a distance of 500 nm from a 50 Pa glycocalyx bound to an endothelium of modulus $E_s$ of about 10 kPa, we find that the glycocalyx properties make the dominant contribution to EHD interactions for $H_0\gtrsim 500$ nm. This suggests that, even under the {\it in vivo} conditions of blood microcirculation, EHD interactions due to wall compliance are entirely controlled by the softness of the $\mu$m-thick glycocalyx.

\end{document}